\documentclass[useAMS,usenatbib]{mn2e}

\usepackage{tabularx}
\usepackage{graphicx}
\usepackage{amsmath}
\usepackage{natbib}

\title[850$\mu$m observations of WISE/radio-selected AGN]{Submillimetre observations of WISE/radio-selected AGN and their environments}

\author[Suzy F. Jones et al.]
{\parbox{\textwidth}{Suzy F. Jones,$^{1}$\thanks{E-mail: sfj8@le.ac.uk}
Andrew W. Blain,$^{1}$
Carol Lonsdale,$^{2}$
James Condon,$^{2}$
Duncan Farrah,$^{3}$
Daniel Stern,$^{4}$
Chao-Wei Tsai,$^{4}$
Roberto J. Assef,$^{5}$
Carrie Bridge,$^{6}$
Amy Kimball,$^{7}$
Mark Lacy,$^{2}$
Peter Eisenhardt,$^{4}$
Jingwen Wu$^{8}$
and Tom Jarrett$^{9}$
}\vspace{0.4cm}\\
\parbox{\textwidth}{$^{1}$University of Leicester, XROA, Department of Physics \& Astronomy, University Road, Leicester LE1 7RH, UK\\
$^2$National Radio Astronomy Observatory, 520 Edgemont Road, Charlottesville, VA 22903-2475 USA\\
$^3$Virginia Polytechnic Institute \& State University, Department of Physics MC 0435, 850 West Campus Drive, Blacksburg, VA 24061, USA\\
$^4$Jet Propulsion Laboratory, California Institute of Technology, 4800 Oak Grove Dr., Pasadena, CA 91109, USA\\
$^5$N\'ucleo de Astronom\'ia de la Facultad de Ingenier\'ia, Universidad Diego Portales, Av. Ej\'ercito Libertador 441, Santiago, Chile\\
$^6$California Institute of Technology MS249-17, Pasadena, CA, 91125, USA\\
$^7$CSIRO Astronomy \& Space Science, Epping, NSW 1710 Australia, PO Box 76\\
$^8$Division of Physics \& Astronomy, University of California Los Angeles, Physics and Astronomy Building, 430 Portola Plaza, Los Angeles, CA 90095-1547, USA\\
$^9$Astronomy Department, University of Cape Town, Rondebosch 7701, Republic of South Africa\\
}}

\begin{document}

\date{Submitted xx/xx/2014}

\pagerange{\pageref{firstpage}--\pageref{lastpage}} \pubyear{2014}

\maketitle

\label{firstpage}

\begin{abstract}
We present JCMT SCUBA-2 850\,$\mu$m submillimetre (submm) observations of 30 mid-infrared (mid-IR) luminous AGN, detected jointly by the WISE all-sky IR survey and the NVSS/FIRST radio survey. These rare sources are selected by their extremely red mid-infrared spectral energy distributions (SEDs) and compact radio counterparts. Further investigations show that they are highly obscured, have abundant warm AGN-heated dust and are thought to be experiencing intense AGN feedback. These galaxies appear to be consistent with an AGN-dominated galaxy, and could be a transient phase of merging galaxies. When comparing the number of submm galaxies (SMGs) detected serendipitously in the surrounding 1.5-arcmin to those in blank-field submm surveys, there is a very significant overdensity, of order 5, but no sign of radial clustering centred at our primary objects. The WISE/radio-selected AGN thus reside in 10-Mpc-scale overdense environments, that could be forming in pre-viralised clusters of galaxies. WISE/radio-selected AGNs appear to be the strongest signposts of high-density regions of active, luminous and dusty galaxies. SCUBA-2 850\,$\mu$m observations indicate that their submm fluxes are low compared to many popular AGN SED templates, hence the WISE/radio-selected AGNs have either less cold and/or more warm dust emission than normally assumed for typical AGN. Most of the targets are not detected, only four targets are detected at SCUBA-2 850\,$\mu$m, and have total IR luminosities $\ge$ 10$^{13}$\,L$_{\odot}$, if their redshifts are consistent with the subset of the 10 SCUBA-2 undetected targets with known redshifts, $z$ $\sim$ 0.44 - 2.86. 
\vspace{0.6cm}
\end{abstract} 


\section{Introduction}

A popular galaxy evolution theory is for major mergers between gas-rich galaxies to cause tidal interactions between the two galaxies and gas is driven into the central regions of the galaxies \citep{sanders88,barnes92,schweizer98,farrah01,veilleux02,hopkins06,hopkins08,volonteri11}. Gas is converted quickly into stars in a period of intense starburst activity ($< 10^{9}$yr), which dominates the luminosity. The active galactic nuclei (AGNs) are fuelled and merge, and with the increased gas supply can accrete at close to or above the Eddington rate \citep{assef14}, dominating the luminosity for a time as a powerful obscured AGN, and then perhaps an optically bright quasar. Feedback from the supermassive blackhole (SMBH) and supernovae can expel all the gas from the galaxy, and quench further star formation and BH accretion, leaving behind a passive elliptical galaxy. Hence, AGN feedback likely plays a crucial role in galaxy evolution. 
There are two modes for AGN feedback: radiative or wind mode, it is intense and short-lived ($\ll$ 10$^8$ years), when the accreting black hole is close to the Eddington limit and affects the distribution of cold gas \citep{farrah12}, and the second is kinetic or maintenance mode, which lasts longer and is less intense than radiative mode, it is jet-driven, and is when the galaxy has a hot halo \citep{fabian12}. From previous observations, radiative mode feedback is important for highly luminous, obscured galaxies, comparable to the galaxies observed in this paper, and kinetic mode feedback is important in nearby massive elliptical galaxies \citep{fabian12}. However, a combination of both radiative and jet-driven feedback modes could be important for the targets of this paper, because the galaxies are extremely luminous and obscured, and also contain radio jets. Observations of AGN feedback, especially at the peak epoch of cosmic star formation, $z$ $\sim$ 2 - 3, are required to fully understand how AGN feedback affects the host galaxy.
%The stellar velocity dispersion, the statistical distribution of orbiting stars in the gravitational potential well (the galactic bulge), has a correlation with the central BH mass of the galaxy, and is known as the \textit{M}-$\sigma$ relationship \citep{ferrarese00,gebhardt00,tremaine02}. This relationship holds for local galaxies, but is currently unclear how it evolves with redshift \citep{yu02,alexander05,borys05,treu07,woo08,bennert11,volonteri11}.  AGN feedback is thought to explain the observed relation \citep{dimatteo05,croton06,hopkins06,yu08,narayanan10,hayward12,ishibashi12}. The \textit{M}-$\sigma$ relationship suggests an evolutionary link between the growth of the BH and the stellar mass of the host galaxy. Observations of AGN feedback, especially at the peak epoch of cosmic star formation, $z$ $\sim$ 2 - 3, are required to fully understand how AGN feedback affects the host galaxy and to see if the \textit{M}-$\sigma$ relationship evolves with redshift.

To observe AGN feedback in action, mid-infrared (mid-IR) selections are very successful \citep{grijp87,low88,spinoglio89,sanders88,keel94,lacy04,stern05,yan05,polletta06,polletta08,gruppioni08,sacchi09,tommasin10,donley12,stern12,assef13,mateos13}. 
NASA's \textit{Wide-field Infrared Survey Explorer} (WISE) \citep{wright10} is able to find luminous, dusty, high-redshift, active galaxies because the hot dust heated by AGN and/or starburst activity can be traced using the WISE 12\,$\mu$m (W3) and 22\,$\mu$m (W4) bands. \citet{eisenhardt12}, \citet{bridge13} and Lonsdale et al. (submitted) have shown that WISE can find different classes of interesting, luminous, high-redshift, dust-obscured AGN. There has been previous work on heavily-obscured, hyper-luminous, WISE-selected AGNs from \citet{eisenhardt12}, \citet{wu12}, \citet{jones14} and \citet{tsai14} who observed galaxies with faint or undetectable flux densities in the 3.4\,$\mu$m (W1) and 4.6\,$\mu$m (W2) bands, and well detected fluxes in the W3 and/or W4 bands, but a radio blind selection. These galaxies, which also host obscured AGNs, are called ``W1W2-dropouts'' or hot, dust-obscured galaxies (Hot DOGs) \citep{wu12,jones14}. Hot DOGs are thought to be consistent with a later, transient phase of a major merger compared to submillimetre galaxies (SMGs) \citep{jones14}, and a comparison of submm observations of the WISE/radio-selected AGNs and their surrounding environments are presented in this paper, to see if they are different. 

The observations of luminous, dusty, high-redshift, active galaxies also revealed significant evidence that the galaxy density in the environments of high-redshift far-IR and mid-IR luminous galaxies appears to be above average \citep{blain04,borys04,scott06,farrah06,gilli07,chapman09,hickox09,cooray10,hickox12}. Clustering of these mid-IR and SMGs could be evidence for massive dark matter halos and highlight bias of this distribution as compared with the underlying dark matter distribution. There is also evidence that SMGs are found in dense environments from \citet{umehata14}, who observed the protocluster SSA22 field with the Astronomical Thermal Emission Camera (AzTEC) on the Atacama Submillimeter Telescope Experiment (ASTE), at 1.1-mm to a depth of 0.7 - 1.3 mJy beam$^{-1}$, and found 10 SMGs correlated with $z = 3.1$ Lyman-alpha emitters (LAEs) in the protocluster. There is also evidence for high-redshift radio galaxies (HzRGs) to reside in overdense regions as traced by dusty galaxies \citep{stevens03,falder10,stevens10,galametz10,galametz12,mayo12,wylezalek13,wylezalek14,hatch14}. This suggests that HzRGs are progenitors of massive galaxies in the centre of rich galaxy clusters in the present-day Universe \citep{stevens03,venemans07,falder10,stevens10,galametz10,galametz12,mayo12,wylezalek13,hatch14}: in particular, the \textit{Herschel} Galaxy Evolution Project (HeRG\'E) found that the HzRG MRC 1138-26 at $z = 2.156$, the Spiderweb galaxy, is a protocluster environment \citep{seymour12}. Further evidence is from the Clusters Around Radio-Loud AGN (CARLA) \textit{Spitzer} programme that looked at the environments of radio-loud AGN (RLAGN) at $1.2 < z < 3.2$, and concluded that RLAGN are in overdense environments in mid-IR wavelengths, and could be signposts of high-redshift galaxy clusters \citep{wylezalek13,hatch14}. \citet{donoso14} looked at $\sim$ 170,000 WISE-selected Hot DOGs and found that obscured AGN are found in denser environments than unobscured AGN. \citet{stevens03} observed seven HzRGs with Submillimetre Common-User Bolometer Array (SCUBA), and compared the number of serendipitous sources in the fields of seven HzRGs and their lambda cold dark matter ($\Lambda$-CDM) simulation predicted distribution (Figure 4), and showed the data and simulations show that either no companions or two companions are found in the HzRG fields. \citet{stevens03} concluded that the radio galaxies had intense, extended star-formation activity and detected on average one serendipitous source per HzRG field.
%High-redshift dusty galaxy populations appear to reside preferentially in clustered and overdense regions of the sky. 

In this paper, we present James Clerk Maxwell Telescope (JCMT) SCUBA-2 \citep{holland13} observations of a subset of WISE/radio-selected AGNs from Lonsdale et al. (submitted). The 30 luminous, dusty, high-redshift, active galaxies are selected from WISE, but also contain compact radio sources, in order to observe AGN feedback and potential overdense environments. They were selected on their red WISE mid-IR colours, with strong compact radio emission in the National Radio Astronomy Observatory (NRAO) Very Large Array (VLA) Sky Survey (NVSS) \citep{condon98} and/or Faint Images of the Radio Sky at Twenty-cm (FIRST) \citep{becker95}. These WISE/radio-selected AGNs are likely to have spectral energy distributions (SEDs) dominated by AGN that could be quenching star formation by the feedback at the highest rate of AGN fuelling, where feedback is most effective and important. The long wavelength SCUBA-2 measurements are needed to understand the cold dust properties of the target and to calculate the total IR luminosity ($L_{IR}$) all the way from 8\,$\mu$m to 1000\,$\mu$m ($L_{IR}$ = $L_{8-1000\mu \textrm{m}}$) (the total IR luminosity).

Section 2 summaries the details of the sample selection from WISE and NVSS/FIRST data (Lonsdale et al. submitted).
Section 3 describes the SCUBA-2 observations. 
Section 4 reports the SCUBA-2 results for SEDs and total IR luminosities ($L_{IR}$), and appropriate templates in fitting the data is discussed. The maximum permitted luminosities of underlying host galaxy components are also calculated. The overdensity of serendipitous SMG sources in the SCUBA-2 fields is determined by comparison with blank-field submm surveys. The overdensity results from Hot DOGs \citep{jones14} are compared, and we compared the distribution of serendipitous SMG sources within each SCUBA-2 field to HzRG fields \citep{stevens03}.

Throughout this paper we assume a $\Lambda$-CDM cosmology with H$_0$ = 71\,km\,s$^{-1}$Mpc$^{-1}$, $\Omega_{\rm{m}}$ = 0.27 and $\Omega_\Lambda$ = 0.73. WISE catalogue magnitudes are converted to flux densities using zero-point values on the Vega system of 306.7, 170.7, 29.04 and 8.284\,Jy for WISE 3.4, 4.6, 12 and 22\,$\mu$m wavelengths, respectively \citep{wright10}.

\section{Sample Selection}

The 30 galaxies observed here with JCMT SCUBA-2 are a subset from the WISE/radio-selected AGNs described in more detail by Lonsdale et al. (submitted), of which 49 Southern galaxies were observed with ALMA in cycle 0. JCMT SCUBA-2 was used to observe galaxies from an independent sample in the Northern hemisphere. JCMT SCUBA-2 observations were also used to compare to the SCUBA-2 observations of Hot DOGs reported by \citet{jones14}, and to observe the potential overdense environments surrounding the WISE/radio-selected AGNs. 

\subsection{WISE}

WISE surveyed the entire sky at wavelengths of 3.4, 4.6, 12 and 22\,$\mu$m (W1-W4) from January 2010 to January 2011 \citep{wright10}. One of the primary science goals was to identify the most luminous galaxies in the observable Universe, which can be accomplished due to WISE obtaining much greater sensitivity than previous all-sky IR survey missions. For example, \textit{IRAS} yielded catalogued source sensitivities of 0.5\,Jy at 12, 25 and 60\,$\mu$m and 1\,Jy at 100\,$\mu$m and angular resolutions that varied from 0.5\,arcmin at 12\,$\mu$m to about 2\,arcmin at 100\,$\mu$m \citep{neugebauer84}; compared to WISE that achieved 5-$\sigma$ source sensitivities better than 0.054, 0.071, 0.73 and 5.0\,mJy and angular resolutions of 6.1, 6.4, 6.5 and 12.0\,arcsec in the W1 to W4 bands, respectively \citep{wright10,jarrett11}. The objects observed here are selected from the WISE AllWISE Source catalog\footnote{http://wise2.ipac.caltech.edu/docs/release/allwise/}, with IR magnitudes derived using point source profile-fitting \citep{cutri12}. 

\subsection{NVSS/FIRST}

NVSS is a 1.4\,GHz continuum survey of the entire sky north of -40$^\circ$ declination, and covers 82\% of the sky, with an angular resolution of 45\,arcsec \citep{condon98}. NVSS ran from 1993 to 1997 and catalogued $\sim$ 1.8 $\times$ 10$^6$ discrete sources with a completeness limit $\ge$ 2.5\,mJy. 

FIRST is a 1.4\,GHz survey over 10,000 degree$^2$ of the North and South Galactic Caps at 20\,cm \citep{becker95}. It produced 5\,arcsec-resolution maps, with a typical root mean square (RMS) noise level of 0.15 mJy. FIRST has a higher resolution than NVSS and so when both were available, FIRST positional data was used. 
% with 1.8\,arcsec pixel maps

The selection criteria $0.1 < \rm{S}_{22\,\mu m} / \rm{S}_{1.4\,GHz} < 1$, ensures that the sample are radio-intermediate, not as radio-bright as standard radio-galaxies, 10$^{40}$ - 10$^{45}$ erg s$^{-1}$ \citep{kellermann74}. The radio data maps the synchrotron emission, and so by using this selection cut ensures that their 850\,$\mu$m fluxes are not contaminated by synchrotron emission. The brightest radio source in this sample has a ${S}_{1.4\,GHz}$ = 250.2\,mJy, and most have a radio flux density of between 10-100\,mJy at 1.4\,GHz. The weighted average ${S}_{1.4\,GHz}$ = 104.8\,mJy, and adopting a power law spectral index of $\alpha = -1.0$, which appears to be appropriate for radio-intermediate targets \citep{fanti00,giacintucci07,varenius14}, the contribution to the 850\,$\mu$m flux would be 0.4\,mJy. When looking at the radio-loudest (S$_{1.4\,\rm{GHz}}$ = 250.2\,mJy) and the radio-quietest (S$_{1.4\,\rm{GHz}}$ = 12.2\,mJy) targets, the contribution to the 850\,$\mu$m flux would be 1.0\,mJy and 0.04\,mJy, respectively. Therefore, the sample's 850\,$\mu$m fluxes, probed to a depth of 2.1\,mJy/beam, should not be contaminated strongly by synchrotron emission. 

The radio fluxes of the sample are similar to the luminous-infrared galaxies (LIRGs)\footnote{Luminous-infrared galaxies (LIRGs), ultra-luminous infrared galaxies (ULIRGs) and hyper-luminous infrared galaxies (HyLIRGs) have characterising total infrared luminosities (8-1000\,$\mu$m) of $L_{8-1000\mu \textrm{m}} > 10^{11}$ L$_\odot$, $L_{8-1000\mu \textrm{m}} > 10^{12}$ L$_\odot$ and $L_{8-1000\mu \textrm{m}} > 10^{13}$ L$_\odot$, respectively \citep{sanders&mirabel96,lonsdale06}},  observed with the Very Long Baseline Array (VLBA) that had strong radio cores and were found to be AGN dominated \citep{lonsdale03}. This could imply that the strong radio emission of the WISE/radio-selected AGNs is evidence of significant non-thermal contributions and could be from the AGN jets. For more details see Lonsdale et al. (submitted).

The NVSS/FIRST catalog was cross-matched with the WISE catalog on $<$ 7\,arcsec scales, which was best for reliability and completeness, and excluding the region within 10 degrees of the Galactic plane to avoid asymptotic giant branch (AGB) stars and saturation artefacts. Matched targets with very red WISE colours were selected, the selection cut $(\rm{W2 - W3}) + 1.25(\rm{W1 - W2}) > 7$, W4 $>$ 10\,mJy and W3 $\ge$\,7mJy were used because coverage levels and sensitivity varies over the sky in the WISE survey (Lonsdale et al. submitted). The NVSS and WISE selection cuts were made to ensure that the targets had steep mid-IR WISE SEDs from W1-W4, which are consistent with AGN SEDs, with compact radio-intermediate structures. The final selection cut was for the targets to have faint or no optical counterparts to the Sloan Digital Sky Survey (SDSS) depth $r$-[24] $>$ 12.3, to avoid confusion with bright or extended low-redshift galaxies. 

These cuts led to 156 selected galaxies . The surface density of WISE/radio-selected high-redshift galaxies over the whole sky, to this magnitude limit, is 0.003 deg$^{-2}$ and points to this population being exceptionally rare and perhaps an interesting transition population. They have no bright optical counterpart and are ultraluminous if at redshift $z$ $>$ 0.5. Ten targets have known redshifts, with a range 0.444 $<$ $z$ $<$ 2.855, and most of the other targets are expected to be in the same range, because they have similar WISE colour cuts. These redshifts ensure that the targets will benefit from the negative K-correction when observed at submm wavelengths.

\section{Observations}

\subsection{JCMT SCUBA-2}

SCUBA-2 is a submm 450/850\,$\mu$m bolometer camera with eight 32 x 40 pixel detector arrays, each with a field-of-view of 2.4\,arcmin$^2$. The diffraction-limited beams have full-width half maxima (FWHM) of approximately 7.5 and 14.5\,arcsec, respectively \citep{holland13}.

From the 130 WISE/radio-selected AGNs, 30 that could be observed in the Northern hemisphere, were observed using SCUBA-2 on the 15-m JCMT atop Mauna Kea in Hawaii, primarily in August 2013 but also on other nights through the 12B semester, from August 2012 to January 2013, and in the 13B semester, from August 2013 to January 2014, and the 14A semester, from February 2014 to September 2014. The optical depth at 225\,GHz, $\tau$$_{225}$, during the observations was in the range of JCMT Band 2 conditions: 0.05 $<$ $\tau$$_{225}$ $<$ 0.08. The corresponding opacities for each atmospheric window, 450\,$\mu$m and 850\,$\mu$m,  were 0.61 $<$ $\tau$$_{450}$ $<$ 1.18 and 0.24 $<$ $\tau$$_{850}$ $<$ 0.40 \citep{dempsey13}. Therefore, we could not use any 450\,$\mu$m data because the atmospheric opacity was too great, and noise levels were too high.

All observations were taken in the ``CV Daisy" observing mode that produces a 12-arcmin diameter map, with the deepest coverage in a central 3-arcmin diameter region \citep{holland13}. The target stays near the centre of the arrays and the telescope performs a pseudo-circular pattern with a radius of 250\,arcsec at a speed of 155\,arcsec\,s$^{-1}$. This mode is best for point-like sources or sources with structure smaller than 3-arcmin scales. Each scan was 25\,mins long and there were three scans per target, totalling a exposure time per target of 75\,mins. 

Pointing checks were taken throughout the nights. The calibration sources observed were Uranus, CRL 2688, CRL 618 and Mars. Calibrations were taken at the start and end of every night in the standard manner \citep{dempsey13}.

\section{Results}

\subsection{Photometry}

The maps were reduced with the STARLINK SubMillimeter User Reduction Facility (SMURF) data reduction package with the ``Blank Field" configuration suitable for low signal-to-noise ratio (SNR) point sources \citep{chapin13}. SMURF performs pre-processing steps to clean the data by modelling each of the contributions to the signal from each bolometer, flatfields the data and removes atmospheric emission, and finally regrids to produce a science-quality image. Using the STARLINK PIpeline for Combining and Analyzing Reduced Data (PICARD) package the maps were mosaiced with all three observations per target, beam-match filtered with a 15\,arcsec FWHM Gaussian and calibrated with the flux conversion factor (FCF) of 2.34\,Jy\,pW$^{-1}$\,arcsec$^{-2}$ (appropriate for aperture photometry) or 537\,Jy\,pW$^{-1}$\,beam$^{-1}$ (in order to measure absolute peak fluxes of discrete sources) that is pertinent for 850\,$\mu$m data \citep{dempsey13}.

\begin{figure*}
\includegraphics[width=11cm,height=11cm]{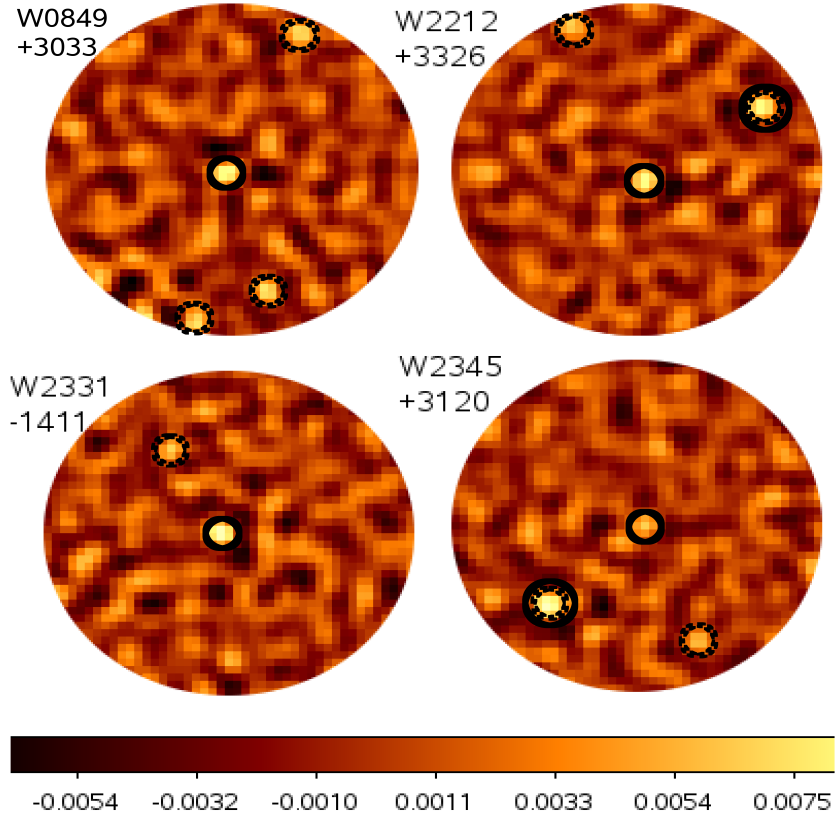}
\caption{SCUBA-2 850\,$\mu$m 1.5\,arcmin radius maps of the four detected targets; W0849$+$3033, W2212$+$3326, W2331$-$1411 and W2345$+$3120. The solid circles show the 15-arcsec beam-sized apertures centred on the WISE positions of the targets. Serendipitous sources brighter than 3\,$\sigma$ and within 1.5\,arcmin of the WISE target are shown by the dotted 15-arcsec beam-sized circles, and serendipitous sources brighter than 4\,$\sigma$ are shown by a dotted 15-arcsec beam-sized circle surrounded by a solid black circle. It is shown that the number of detected serendipitous SMG sources is independent of the detection of the WISE/radio-selected AGN target, when compared to Figure~\ref{undet}. The colour flux bar at the bottom is in Jy. North is up, East is to the left.}
\label{det}
\end{figure*}

\begin{figure*}
\includegraphics[width=11cm,height=11cm]{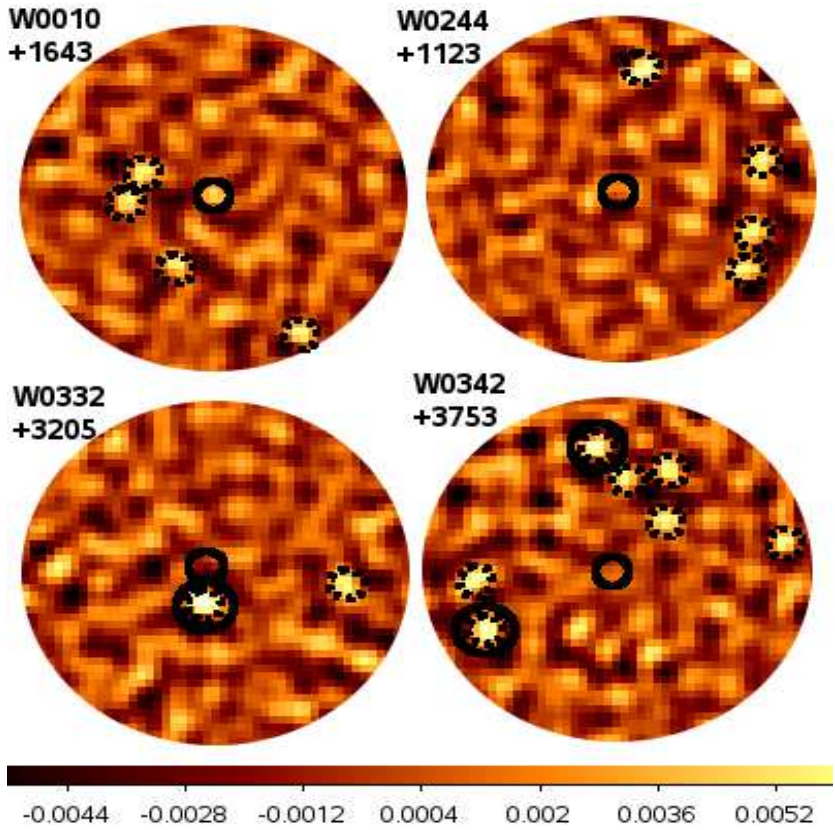}
\caption{SCUBA-2 850\,$\mu$m 1.5\,arcmin radius maps of a sample of 4 undetected targets ordered by RA DEC: W0010$+$1643, W0244$+$1123, W0332$+$3205 and W0342$+$3753. The solid circles show the 15-arcsec beam-sized apertures centred on the WISE positions of the targets. Serendipitous sources brighter than 3\,$\sigma$ and within 1.5\,arcmin of the WISE target are shown by the dotted 15-arcsec beam-sized circles, and serendipitous sources brighter than 4\,$\sigma$ are shown by the dotted 15-arcsec beam-sized circle surrounded by a solid black circle. It is shown that the number of detected serendipitous SMG sources is independent of the detection of the WISE/radio-selected AGN target, when compared to Figure~\ref{det}. The colour flux bar at the bottom is in Jy. North is up, East is to the left.}
\label{undet}
\end{figure*}

\begin{figure*}
\begin{centering}
\includegraphics[width=15cm,height=7cm]{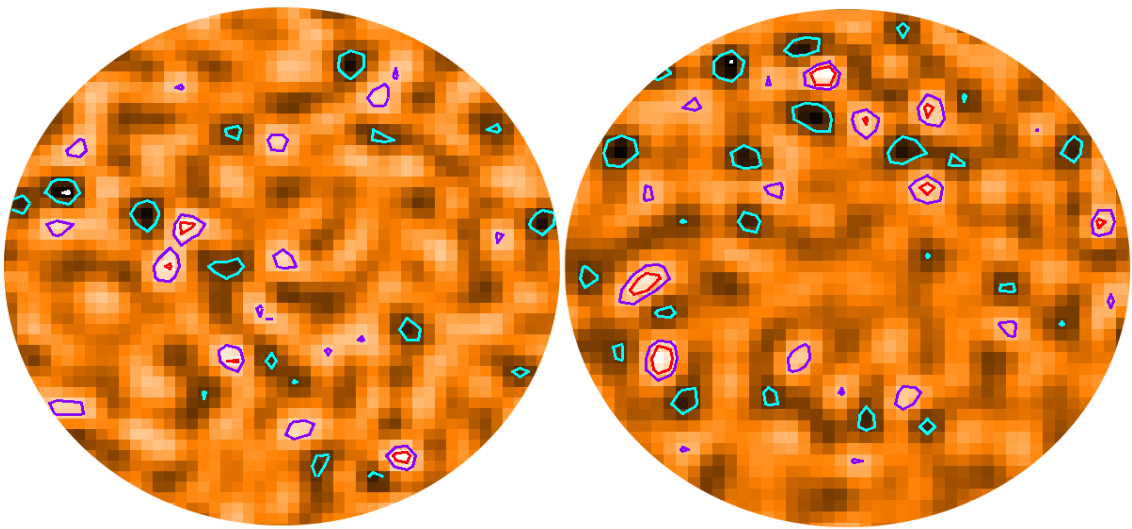}
\caption{SCUBA-2 850\,$\mu$m 1.5\,arcmin radius maps showing the central region of the first target when ordered by RA DEC W0010$+$1643 and the field with the largest number of detected serendipitous SMG sources (7) W0342$+$3753, with flux density contours: white represents negative 3\,$\sigma$, blue represents negative 2\,$\sigma$, purple represents positive 2\,$\sigma$ and red represents positive 3\,$\sigma$. North is up, East is to the left.}
\label{contours}
\end{centering}
\end{figure*}

The 850\,$\mu$m flux densities of the 30 WISE/radio-selected AGNs at their WISE positions and the noise level in the maps are presented in Table 1, with typical RMS noise of 2.1 mJy/beam. Four are detected at greater than 3\,$\sigma$ significance; 21 targets had less significant positive flux measurements at the WISE position, with a typically significance 1.4\,$\sigma$; and five targets had negative flux measurements at the WISE position. Flux density limits were measured in an aperture diameter of 15\,arcsec, the FWHM size of the telescope beam. This was an appropriate aperture size, because for the detected sources, the measured flux densities with a 15\,arcsec aperture size, are consistent with the measured peak flux densities. These results can be compared to Lonsdale et al. (submitted) who observed 49 WISE/radio-selected AGNs in a snapshot mode with the Atacama Large Millimeter/submillimeter Array (ALMA). They detected a larger fraction of targets, with 27 out of 49 sources detected at greater than 3\,$\sigma$ significance, in substantially deeper observations, with RMS noise of 0.3-0.6 mJy. Figure~\ref{det} and Figure~\ref{undet} show the most sensitive 3-arcmin diameter SCUBA-2 850\,$\mu$m DAISY fields of the four detected and four undetected WISE/radio-selected AGNs ordered in ascending RA, respectively. It is shown that the number of detected serendipitous SMG sources is independent of the detection of the WISE/radio-selected AGN target, when comparing Figure~\ref{det} to Figure~\ref{undet}. The typical offset of the WISE position compared with the SCUBA-2 position of the detected targets was small: 2\,arcsec. Figure~\ref{contours} shows the SCUBA-2 850\,$\mu$m 1.5\,arcmin map of the targets W0010$+$1643 and W0342$+$3753, with contours representing 2 and 3\,$\sigma$ positive and negative sources.

\begin{figure}
\begin{centering}
\includegraphics[width=7cm,height=7cm]{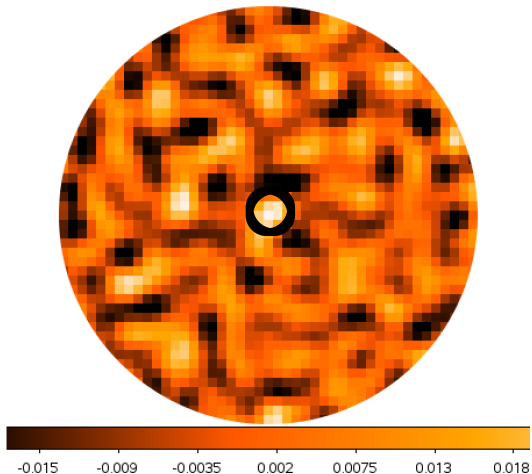}
\caption{SCUBA-2 850\,$\mu$m 1.5\,arcmin radius map showing the central regions of the maps of the 26 undetected targets stacked together. The circle shows the central 15-arcsec area. The colour flux bar at the bottom is in Jy. North is up, East is to the left.}
\label{stack}
\end{centering}
\end{figure}

When the images of the 26 undetected sources are stacked together into one image (Figure~\ref{stack}), centred on the WISE-determined position of each targets, the net flux is 32.7 $\pm$ 9.0\,mJy in the central 15\,-arcsec aperture, a detection of 3.6\,$\sigma$. The typical flux of an undetected target is thus likely to be approximately 1.8\,mJy, which is comparable to the ALMA snapshot detection limit of Lonsdale et al. (submitted) who found a typical flux limit for an undetected target to be 1.2\,mJy. The 26 undetected targets are consistent with being on average 4.5 times fainter than the four detected targets; this is consistent with Hot DOGs from \citet{jones14}. To get deeper observations with SCUBA-2, to be able to detect more of the targets at 3\,$\sigma$, would require several more hours of integration per target, beyond the existing 75\,mins, but would not add much more value to this stacked result, and the SCUBA-2 confusion limit is $\sim$1\,mJy\,beam$^{-1}$ \citep{geach13}.

To test whether the positive flux density of the 26 targets with upper limits is likely to be real, 60 random points were sampled from the maps, and the stacked average flux density was 0.0 $\pm$ 0.3\,mJy. This is consistent with the positive flux densities from the WISE/radio-selected AGNs with upper limits being due to significantly fainter, individually undetected targets. 

%W1107$+$3421, W1212$+$4659, W1409$+$1732, W1428$+$1113, W1501$+$1324, W1517$+$3523, W1630$+$5126, W1703$+$2615, W1717$+$5313, W2126$-$0103, W2133$-$1419, W2212$-$1253, W2222$+$0951, W2226$+$0025, W2230$-$0720, W2325$-$0429

\subsection{SEDs}

\subsubsection{WISE-derived SEDs}

\begin{figure}
\includegraphics[width=84mm]{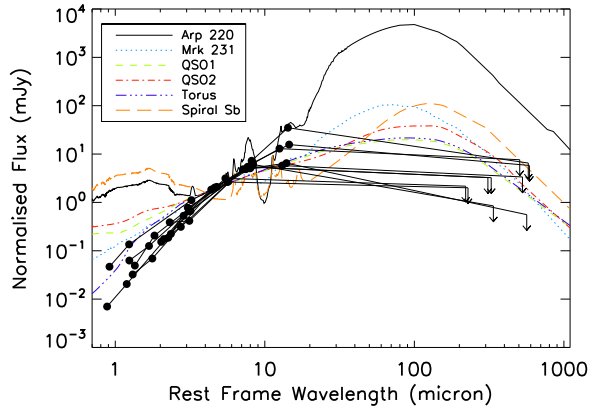}
\caption{SEDs of the 10 WISE/radio-selected AGNs with known redshifts, including the 850\,$\mu$m SCUBA-2 data in rest-frame wavelengths with Arp 220, Mrk 231, QSO 1, QSO 2, Torus and Spiral Sb galaxy templates from \citet{polletta07}, normalised at rest frame 3 $\mu$m. Detections are represented by filled circles, while 2\,$\sigma$ upper limits are represented by arrows. The data points for the WISE/radio-selected AGNs are connected for clarity, and do not represent the true SED. The 10 WISE/radio-selected AGNs are not well fitted by the standard AGN templates; with extra dust extinction required, which is similar to the Hot DOG SEDs \citep{jones14}.}
\label{sedall}
\end{figure}

\begin{figure}
\includegraphics[width=84mm]{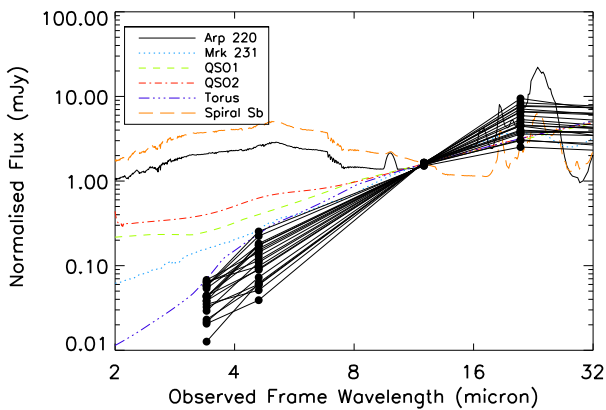}
\caption{Mid-IR SEDs of the 20 WISE/radio-selected AGNs showing the range of WISE data, in observed-frame wavelengths Arp 220, Mrk 231, QSO 1, QSO 2, Torus and Spiral Sb galaxy templates \citep{polletta07} redshifted to $z$ $=$ 2  and normalised at observed frame 12 $\mu$m. Detections are represented by filled circles, while 2\,$\sigma$ upper limits are represented by arrows. The data points for the WISE/radio-selected AGNs are connected for clarity, and do not represent the true SED.}
\label{sedallnoz}
\end{figure}

The SEDs of the 10 SCUBA-2 WISE/radio-selected AGNs with known redshifts are shown in Figure~\ref{sedall}, normalised at rest-frame 3\,$\mu$m and shown at rest-frame wavelengths in order to compare with various galaxy SED templates \citep{polletta07} in order to illustrate the nature of the WISE/radio-selected AGNs. The mid-IR SEDs of the 20 WISE/radio-selected AGNs with unknown redshifts are shown in Figure~\ref{sedallnoz}, with the Polletta SED templates redshifted to $z = 2$. The Polletta galaxy templates are Arp 220 (starburst-dominated galaxy), Mrk 231 (heavily obscured AGN-starburst composite), QSO 1 and 2 (optically-selected QSOs of Type 1 and 2), spiral (Spiral Sb), torus (type-2 heavily obscured QSO, that is believed to be an accreting SMBH with a hot accretion disk surrounded by dust and Compton-thick gas in a toroidal structure \citep{krolik88}). The SEDs are broadly similar at mid-IR wavelengths. They have a steep red power-law IR (1-5\,$\mu$m) section, and a potential mid-IR peak from hot dust emission that is bluewards of the AGN SED templates, turning over to a Rayleigh-Jeans spectrum longwards of 200\,$\mu$m due to the coolest dust emission. The SEDs are not well-represented by any of the Polletta templates; the closest fitting template is the single Polletta torus template. The SEDs are consistent with those of the 49 ALMA observed WISE/radio-selected AGNs (Lonsdale et al. submitted). All WISE/radio-selected AGNs have hotter WISE/submm colours than standard pre-existing AGN template SEDs.

The Hot DOGs \citep{wu12,jones14} have broadly similar SED shapes compared to the WISE/radio-selected AGNs here and from Lonsdale et al. (submitted): they all have bluer mid-IR sections and turn over into the submm at shorter wavelengths than typical SED templates. However, the Hot DOGs have steeper mid-IR sections that could be because these galaxies have redder WISE colour selection cuts and a greater average redshift, for example the average redshift in this paper is $z$ = 1.3, and in Lonsdale et al. (submitted) the average redshift is $z$ = 1.7, whereas the average redshift in \citet{jones14} was higher at $z$ = 2.7. The WISE colours for the WISE/radio-selected AGNs are (W1 $-$ W2) = 1.80 and (W2 $-$ W3) = 4.80, compared with Hot DOGs, (W1 $-$ W2) = 1.77 and (W2 $-$ W3) = 6.13. When K-correcting the Hot DOG WISE colours to WISE/radio-selected AGN redshifts, the WISE colours expected are (W1 $-$ W2) = 3.56 and (W2 $-$ W3) = 6.38. The WISE colours show that the Hot DOGs appear to be mid-IR redder, and could be due the sources having more obscuration than the WISE/radio-selected AGNs, which can be seen in the SEDs in Figure~\ref{sedall} when compared to the SEDs in Figure 5 from \citet{jones14}. Alternatively, it could be because the WISE-selected Hot DOGs are typically at a higher redshift, and were selected to be mid-IR redder.

%and the dusty torus template from \citet{rowan08} and three clumpy torus models (Torus 1 N08 with viewing angle i=0, opening angle $\sigma$=30, ratio of the external to internal radii Y=100, number of clouds N$_0$=3, power-law index of radial distribution q=1 and optical depth $\tau$$_\nu$=20, Torus 2 N08 with i=0, $\sigma$=60, Y=30, N$_0$=3, q=1 and $\tau$$_\nu$=20, Torus 3 N08 with i=20, $\sigma$=60, Y=100, N$_0$=9, q=0 and $\tau$$_\nu$=20) from \citet{nenkova08} The torus models are a smooth dusty torus template from \citet{rowan08} and three clumpy torus models from \citet{nenkova08}; the first torus model with viewing angle i=0, opening angle $\sigma$=30, ratio of the external to internal radii Y=100, number of clouds N$_0$=3, power-law index of radial distribution q=1 and optical depth $\tau$$_\nu$=20, the second torus model with i=0, $\sigma$=60, Y=30, N$_0$=3, q=1 and $\tau$$_\nu$=20, and third torus model with i=20, $\sigma$=60, Y=100, N$_0$=9, q=0 and $\tau$$_\nu$=20: there are over 100,000 clumpy torus models, and these three were chosen because they best fit the mid-IR data of luminous QSOs selected from WISE, SDSS and UKIRT Infrared Deep Sky Survey (UKIDSS) \citep{roseboom13}.The torus models from \citet{nenkova08} do not fit the WISE data well however, the \citet{rowan08} model with assumption of smooth dust is a better representation. 

\subsubsection{SCUBA-2-derived SEDs}

\begin{figure}
\includegraphics[width=84mm]{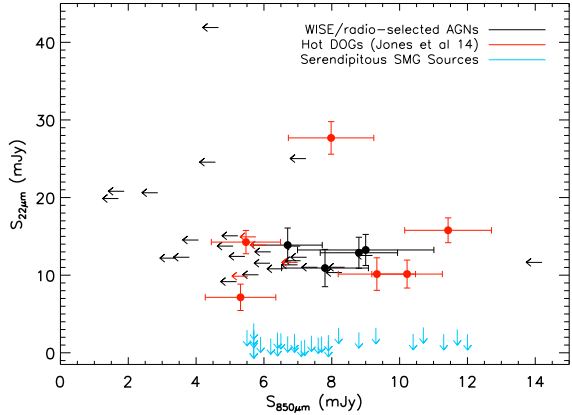}
\caption{The W4 (22\,$\mu$m) flux density versus 850\,$\mu$m flux density for the 30 WISE/radio-selected AGNs in black, the 10 Hot DOGs in red \citep{jones14}, and the 39 serendipitous SMG sources detected around the WISE/radio-selected AGNs that have WISE data in blue. Detections are represented by filled circles, while 2\,$\sigma$ upper limits are represented by arrows.}
\label{hist85022}
\end{figure}

Normalised to the WISE data at rest-frame 3\,$\mu$m that lies within the WISE rest-frame wavelength range for all of our targets, the SCUBA-2 data in the normalised SED shows that the detected WISE/radio-selected AGNs have less submm emission than the Polletta torus template, with an average submm flux three times less than the template, Figure~\ref{sedall}. Most of the undetected WISE/radio-selected AGNs have limits that demand their submm emission to be less than the Polletta torus template, with a flux difference factor range of 1-9. Less submm emission is also seen in the SEDs of Hot DOGs \citep{wu12,jones14}. This leads to the suggestion that the WISE/radio-selected AGNs have less cold dust in the host galaxy and/or on the outer edge of the torus, and hence the torus could be denser and smaller than assumed in the template. Alternatively, less submm emission could be due to excess mid-IR emission from AGN heating relative to the torus template \citep{wu12}. The 22\,$\mu$m (W4) flux densities and the 850\,$\mu$m flux densities of the 30 WISE/radio-selected AGNs and the 10 Hot DOGs are comparable, as shown in Figure~\ref{hist85022}: the serendipitous SMG sources detected around the WISE/radio-selected AGNs that have WISE data are also plotted in blue and will be discussed in Section 6.2. The Hot DOGs in \citet{jones14} are submm brighter, because 60\,$\%$ Hot DOGs are submm detected compared to 13\,$\%$ WISE/radio-selected AGNs; the average submm flux of detected Hot DOGs was 8.3 $\pm$ 1.8 \,mJy, which is similar to the average submm flux of detected WISE/radio-selected AGNs, 8.1 $\pm$ 2.2\,mJy. However, the Hot DOGs are typically at $z \sim 2.7$, compared to $z \sim 1.7$ for the WISE/radio-selected AGNs.  Considering that when K-correcting the WISE/radio-selected AGNs 850\,$\mu$m flux density to the Hot DOG redshift, it should be S$_{850 \mu \textrm{rm}}$ 16 $\pm$ 3 \,mJy, indicating the WISE/radio-selected AGNs are submm fainter than the Hot DOGs. However, this is only an estimate because there are only a small number (10) known redshifts.

\begin{figure}
\includegraphics[width=84mm]{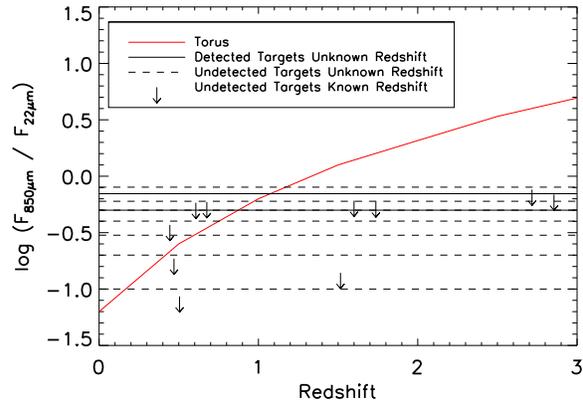}
\caption{The submm to mid-IR ratio (F$_{850 \mu \textrm{m}}$ / F$_{22 \mu \textrm{m}}$) of the 10 targets with known redshifts. The detected targets with unknown redshifts have solid lines, and the undetected targets have dashed lines across the redshift range 0.3 $\le$ $z$ $\le$ 3. The red solid line represents the Polletta torus template \citep{polletta07} as a function of redshift, and 2\,$\sigma$ upper limits are represented by arrows.}
\label{85022}
\end{figure}

To investigate if the SEDs are dominated by warm or cool dust the submm to mid-IR ratios (F$_{850 \mu \textrm{m}}$ / F$_{22 \mu \textrm{m}}$) of the 30 targets in the observed frame are listed in Table 1; for undetected targets, the 850\,$\mu$m 2\,$\sigma$ upper limits are reported. The weighted average F$_{850 \mu \textrm{m}}$ / F$_{22 \mu \textrm{m}}$ of the four detected targets is 0.7 $\pm$ 0.2, where the error is the weighted standard error. This is consistent with the weighted average F$_{850 \mu \textrm{m}}$ / F$_{22 \mu \textrm{m}}$ = 0.6 $\pm$ 0.1 of Hot DOGs \citep{jones14}. Figure~\ref{85022} shows these ratios in the observed sample and for the Polletta AGN torus templates as a function of redshift. Most of the targets have a lower submm to mid-IR flux ratio than the torus template (Figure~\ref{85022}), but show no sign of an expected K-correction with redshift as expected from the template SEDs: in particular the five highest redshift WISE/radio-selected AGNs have submm emission that lies beneath the torus template, and similar observed submm to mid-IR ratios to the lower redshift WISE/radio-selected AGNs. They also have the most luminous luminosities dominated by the mid-IR emission, which could suggest that they have hotter effective dust temperatures compared with the rest of the sample. This is in agreement with the results from submm observations of Hot DOGs \citep{wu12,jones14}, and suggest that Hot DOGs and WISE/radio-selected AGNs are both dominated by warm dust, with lower F$_{850 \mu \textrm{m}}$ / F$_{22 \mu \textrm{m}}$ than standard torus SED templates.

\subsection{Luminosities}

A conservative lower limit to the total IR luminosities of the galaxies were estimated by connecting all the WISE and SCUBA-2 data points with power-laws and then integrating, without extrapolating beyond the range of the data in wavelength, ($L_{8\mu \textrm{m}-\textrm{SCUBA2}}$), the targets without redshift have assumed redshifts of z$=$2.0. This luminosity is a very conservative estimate because any strong peak in the SED would not be included in the power-law interpolation, and it spans a minimal wavelength range. The resulting $L_{8\mu \textrm{m}-\textrm{SCUBA2}}$ values for the 30 WISE/radio-selected AGNs are presented in Table 2. The four detected targets have a range from (1.7 $\pm$ 2.5) $\times$ 10$^{13}$\,L$_{\odot}$ to (6.2 $\pm$ 1.7) $\times$ 10$^{13}$\,L$_{\odot}$, if their redshifts are all consistent with the subset of 10 targets with known redshifts $z$ $\sim$ 0.4 - 2.9, using $z = 2$. This classifies them as hyper-luminous infrared galaxies (HyLIRGs)$^{2}$.

%Exceptionally bright galaxies can often be found to be gravitational lensed, for example \citet{eisenhardt96,williams96,solomon05,vieira10,negrello10,bussmann13}. However, the WISE/radio-selected AGN luminosities are thought to be intrinsic and not due to gravitational lensing: high-resolution imaging programmes (Bridge et al. in prep.; Petty et al. in prep) from \textit{Hubble Space Telescope} (\textit{HST}) and ground-based telescopes of a subset of WISE/radio-selected AGNs show no obvious lensed structures \citep{wu14}. Resolved near-IR \textit{HST} observations show the population to have a range of morphologies from clumpy and extended to point-like (Bridge et al. in prep.; Petty et al. in prep.). This suggests that the WISE/radio-selected AGNs are indeed amongst the most intrinsically luminous galaxies in the universe (Eisenhardt et al. 2012; Tsai et al. in prep.). 

We also use the SCUBA-2 data to limit the luminosity of an underlying extended host galaxy not associated with the WISE/radio-selected AGN, powered by star-forming activity. A spiral (Sb) galaxy SED template and a warmer ULIRG/starburst-type (Arp 220) SED template were normalised to account for all of the SCUBA-2 850\,$\mu$m flux density, and then the maximum total luminosity of these components of the SED were estimated by integrating under the Sb or Arp 220 template. This approach assumes that an underlying extended dusty galaxy, disconnected from the mid-IR emission of the AGN, accounts for all of the measured SCUBA-2 flux. It is assumed to be solely due to star-formation rather than an AGN. An Sb host galaxy template cannot exceed $\sim$ 4\,$\%$ of the inferred WISE/radio-selected AGN luminosity from the four detected targets. This would give a Sb luminosity of 4.9 $\times$ 10$^{12}$\,L$_{\odot}$, 80 times more luminous than the Milky Way, with an equivalent star formation rate (SFR) of $\sim$ 110\,M$_\odot$\,yr$^{-1}$. An Arp 220 ULIRG template that accounts for all the 850\,$\mu$m flux has a luminosity of 1.1 $\times$ 10$^{13}$\,L$_{\odot}$, $\sim$ 10\,$\%$ of the inferred WISE/radio-selected AGN luminosity from the four detected targets, with an equivalent SFR of $\sim$ 250\,M$_\odot$\,yr$^{-1}$. This emphasises that the full WISE/radio-selected AGN SED from 8-1000\,$\mu$m has a small contribution from cold far-IR dust traced by the SCUBA-2 observations, and is dominated by hot dust and mid-IR emission from the AGN. This is consistent with the results from SCUBA-2 observations of Hot DOGs, which found that an underlying Sb-type galaxy SED template contributes less than 2\,$\%$ to the typical total Hot DOG IR luminosity; however, an Arp 220-type galaxy SED template contributes less than 55\,$\%$, this is because the Hot DOGs are submm brighter and could suggest there is more star-formation activity in Hot DOGs compared to WISE/radio-selected AGNs.
%(6 $\times$ 10$^{10}$\,L$_{\odot}$)

\section{Environments Around WISE/radio-selected AGNs}

Serendipitous SMG sources are detected in the deepest SCUBA-2 1.5-arcmin-radius regions around the WISE/radio-selected AGNs. To investigate if there is an overdensity of SMGs in the WISE/radio-selected AGN fields, the serendipitous source number counts are compared with those in two different blank-field submm surveys, in a similar method to \citet{jones14}. To provide another test, 1.5-arcmin-radius circles were placed at random in a blank-field submm survey and the number of galaxies counted, again in a similar method to \citet{jones14}. The differential number counts are then measured and compared to three different blank-field submm surveys.

Eighty-one serendipitous 850\,$\mu$m sources were detected at greater than 3\,$\sigma$ in the 30 SCUBA-2 maps, and 11 sources were detected at greater than 4\,$\sigma$; see Table 3. The total area surveyed was 212\,arcmin$^2$, or about 4000 SCUBA-2 850\,$\mu$m beams.  
%the 3$\sigma$ to 4$\sigma$ ratio is $\sim$ 7
There are 8$\pm$3 negative peaks in the 30 maps above the same 3\,$\sigma$ threshold (see Table 3), consistent with the six 3\,$\sigma$ negative peaks expected from Gaussian noise, with width of the SCUBA-2 FWHM beam (14\,arcsec).

To see if there is evidence for an overdensity of SMGs in the 30 WISE/radio-selected AGNs fields, the number of serendipitous sources can be compared with the results of blank-field submm surveys. In the LABOCA ECDFS Submm Survey (LESS) survey, \citet{weiss09} detected 126 SMGs in a uniform area of 1260\,arcmin$^2$ surveyed with a noise level of 1.2\,mJy at 870\,$\mu$m. They found evidence for an angular clustering signal on scales of 1\,arcmin. There are 60 LESS sources brighter than our average 3\,$\sigma$ flux density limit of 6.2\,mJy. If the average surface number density of SMGs holds in our SCUBA2 maps, so no cosmic variance between the fields, 10.1 serendipitous sources would be expected within 1.5\,arcmin of our 30 targeted fields; however, we have identified 64. This indicates a relative overdensity of SMGs in our WISE/radio-selected AGN fields by a factor of 6.3 $\pm$ 1.1. 

The average noise level of our maps is 2.1 beam$^{-1}$, higher than the average noise level of the deeper observations of Hot DOGs 1.8 mJy beam$^{-1}$ reported in \citet{jones14}. 

In order to check the effect of our range of sensitivity in each field we also compare the number of SMGs at our average 4\,$\sigma$ noise level with the LESS survey. The number of LESS sources brighter than our average 4\,$\sigma$ flux density limit of 8.2\,mJy is 23 SMGs, which suggests that 3.9 serendipitous sources would be expected in our 30 SCUBA-2 fields. We identified 19 and thus a relative overdensity of SMGs by a factor of 4.9 $\pm$ 1.5. The overdensity above the average 3\,$\sigma$ noise level is consistent with that above the average 4\,$\sigma$ noise level; therefore, the $\sim$3 0\,$\%$ range in the 30-map noise levels is unlikely to have a large effect on the overdensity factor.
%, this is because the WISE/radio-selected AGNs had three DAISY scans per target, compared to four scans per target for the Hot DOGs

A complementary way to test whether there is an overdensity of SMGs near the WISE/radio-selected AGN targets is to place 1.5-arcmin-radius SCUBA-2 field circles at random locations in the LESS field and count the number of sources from the catalogue source positions that would have been detected in our survey, taking into account the differences in depth, by employing  our average 3\,$\sigma$ flux density limit of 6.2\,mJy. In 100 randomly selected 1.5-arcmin-radius circles within LESS, we found 32 $\pm$ 3 LESS sources brighter than 6.2\,mJy; the error is estimated from repeating this process three times. The total number of 1.5-arcmin-radius fields within LESS is $\sim$ 200. Thus, in the 30 WISE/radio-selected AGN maps we would expect 10.1 serendipitous sources based on LESS taking into account our field depths; however, we find 64. There is thus a relative overdensity of SMGs around WISE/radio-selected AGNs by a consistent factor of 6.3 $\pm$ 1.4 as compared with this blank-field. 

In 100 randomly selected 1.5-arcmin-radius circles within LESS, we found 13 $\pm$ 2 LESS sources brighter than the 8.2\,mJy average 4\,$\sigma$ noise level. From this it can be predicted that in the 30 WISE/radio-selected AGN maps there should be 3.9 serendipitous sources; however, we find 19. Thus the relative overdensity of 4\,$\sigma$ SMGs around WISE/radio-selected AGNs is by a factor of 4.9 $\pm$ 1.8 as compared with this blank-field. Again the overdensity using the average 3\,$\sigma$ noise level is consistent with that using the average 4\,$\sigma$ noise level; therefore, the difference in the 30 map noise levels appears not to have a large effect on the overdensity factor.

The overdensity factor of SMGs from the WISE/radio-selected AGN fields to the LESS field ranges from $4.9 - 6.3$, with a weighted average of 5.3. \citet{weiss09} concludes that in the LESS field there is a cumulative source count lower by a factor $\sim$ 2 compared with another deep submm survey, the SCUBA Half-Degree Extragalactic Survey (SHADES) \citep{coppin06}. This implies that our WISE/radio-selected AGN fields could have an overdensity that is tripled compared with other previous submm surveys. 

\begin{figure}
\includegraphics[width=84mm]{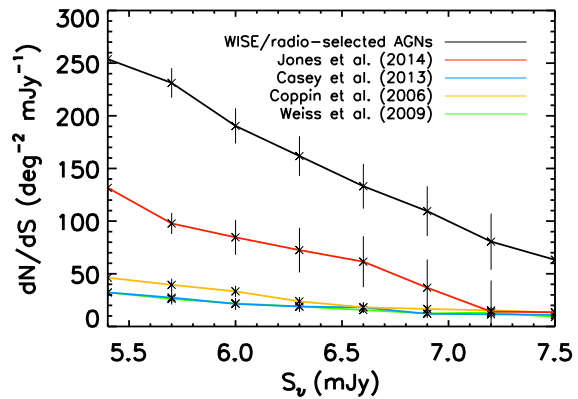}
\caption{The differential number counts for all 30 WISE/radio-selected AGN SCUBA-2 maps, compared with 10 Hot DOG SCUBA-2 maps \citep{jones14}, the LESS survey \citep{weiss09}, the COSMOS survey \citep{casey13} and SHADES \citep{coppin06}.}
\label{nsflux}
\end{figure}

Figure~\ref{nsflux} compares the differential number counts for all 30 WISE/radio-selected AGN SCUBA-2 fields compared to previous blank field submm surveys, including LESS and SHADES, and all 10 Hot DOG SCUBA-2 maps \citep{jones14}. It is seen that the fields around WISE/radio-selected AGNs are overdense compared with other submm surveys, and Hot DOGs are also overdense compared with other submm surveys, which is consistent with the results above. The WISE/radio-selected AGNs are at high Galactic latitudes, and therefore there is less possibility for the high surface number density to be just foreground contamination.

We repeat this approach in another submm blank-field survey. \citet{casey13} used SCUBA-2 to observe the Cosmological Evolution Survey (COSMOS) field over a uniform area of 394\,arcmin$^2$ to a noise level of 0.80\,mJy at 850\,$\mu$m and detected 99 SMGs brighter than 3.6\,$\sigma$. There are 21 COSMOS sources brighter than our average detection threshold of 6.2\,mJy (3\,$\sigma$), which would imply 11.3 serendipitous sources expected in the 30 SCUBA-2 WISE/radio-selected AGN fields. We find 64, implying a relative overdensity of SMGs by a factor of 5.7 $\pm$ 1.4, consistent with 6.3 $\pm$ 1.1 from LESS. Nine COSMOS sources are brighter than our 4\,$\sigma$ flux density limit of 8.2\,mJy, which implies that 4.8 4\,$\sigma$ serendipitous sources would be expected in our 30 SCUBA-2 fields. However, we find 19 and thus a relative overdensity of SMGs by a factor of 4.0 $\pm$ 1.6; consistent with 4.9 $\pm$ 1.5 from LESS. Again, the variation in noise levels of the 30 WISE/radio-selected AGN field does not appear to have a large effect on the overdensity factor. Random 1.5-arcmin-radius circles were not placed in the smaller COSMOS field as only $\sim$ 50 1.5-arcmin-radius fields are available. The COSMOS field is noisier at the edge of the map due to the JCMT “PONG” observing mode, and the noise level could be double at the edge of their 850\,$\mu$m map. Therefore, to detect a source brighter than their 3.6\,$\sigma$ limit, the source would need a flux density of greater than 5.8\,mJy. This could be a problem because our source detection limit is lower at 3\,$\sigma$ and sources could be missed. However, it will not affect our overdensity results because our 3\,$\sigma$ average detection threshold is greater (6.2\,mJy) than their noisiest 3.6\,$\sigma$ limit (5.8mJy), and thus no COSMOS sources would be missed when comparing number counts from their catalogue.

Seventeen serendipitous sources were found in the 10 SCUBA-2 fields of Hot DOGs reported by \citet{jones14}, who concluded that these fields had a SMG overdensity by factor of $\sim$ 2 - 3 compared with the LESS and COSMOS fields.  At our average 3\,$\sigma$ flux density of 6.2\,mJy, there were 9 serendipitous sources found in the 10 SCUBA-2 Hot DOGs, which would imply 26.9 serendipitous sources are expected in these 30 SCUBA-2 WISE/radio-selected AGNs. However, we find 64 and suggests that the WISE/radio-selected AGN fields have a higher density of SMGs than the Hot DOGs by a factor of 2.4 $\pm$ 0.9, and have an even higher overdensity compared with the LESS and COSMOS fields than the Hot DOGs. This is in agreement with \citet{assef14} who found that Hot DOGs reside in overdense environments, as large as the CARLA clusters, but the angular clustering of the objects in the Hot DOG fields were smaller than in the CARLA fields.

\begin{figure}
\includegraphics[width=84mm]{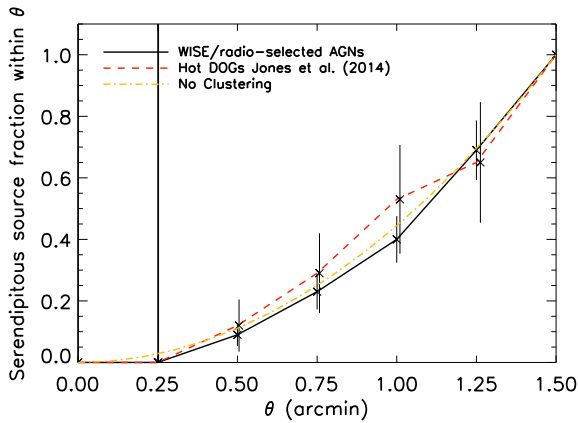}
\caption{The cumulative fraction of the total number of serendipitous sources in each field within different radii of the WISE targets. The solid line shows the fields of the 30 WISE/radio-selected AGNs presented here. The dashed red line shows the 10 Hot DOGs from \citet{jones14}. The yellow dashed-dotted line shows the expected number of serendipitous sources if they are randomly located with no angular clustering. The beam size of SCUBA-2 at 850\,$\mu$m is 14.5\,arcsec; serendipitous sources cannot be detected within the beam.}
\label{clustering_his}
\end{figure}

Figure~\ref{clustering_his} shows the fraction of the total number of serendipitous sources for the 30 SCUBA-2 maps found within cumulative circles of 0.25, 0.5, 0.75, 1.0, 1.25 and 1.5\,arcmin from the WISE target. The expected fraction of the total number of serendipitous sources with no angular clustering, and the result from \citet{jones14} for 10 Hot DOGs that have no evidence of angular clustering on arcmin scales, is also plotted on Figure~\ref{clustering_his}. There is no hint of angular clustering of serendipitous sources around the WISE/radio-selected AGNs on scales less than 1.5\,arcmin, despite the significantly greater average density of submm sources in the WISE fields as compared with blank-field surveys. Clustering of SMGs on larger scales than the extent of the SCUBA-2 field, could be expected because there is tentative evidence of clustering from previous submm studies on scales up to $\sim$8\,arcmin \citep{scoville00,blain04,greve04,farrah06,ivison07,weiss09,cooray10,scott10,hickox12}. Nevertheless, the lack of a clear two-point correlation signal is interesting, because SMG clustering observations can constrain the nature of the host halos around SMGs \citep{cooray10}.

The average separation of a serendipitous source from the WISE central target is 58\,arcsec, which is similar to the Hot DOG serendipitous sources, with a average separation from the WISE central target of 59\,arcsec. This is consistent with Monte Carlo simulations of 1,000 randomly placed serendipitous SMG sources from a central target, where the average separation was 60\,arcsec. When looking at the separations between the serendipitous sources themselves, the average separation is 76\,arcsec for WISE/radio-selected AGN serendipitous sources, 67\,arcsec for Hot DOG serendipitous sources, and 80\,arcsec in 1,000 randomly placed serendipitous SMG sources from Monte Carlo simulations. This would suggest that the serendipitous sources are not angularly clustered on these 1.5\,arcmin scales.

%The purple dashed-dotted-dotted-dotted line shows the Monte Carlo simulation of 1,000 randomly placed serendipitous SMG sources from the central WISE target. and the Monte Carlo simulation of 1,000 randomly placed serendipitous SMG sources on a SCUBA-2 field from the central WISE target,
\begin{figure}
\includegraphics[width=84mm]{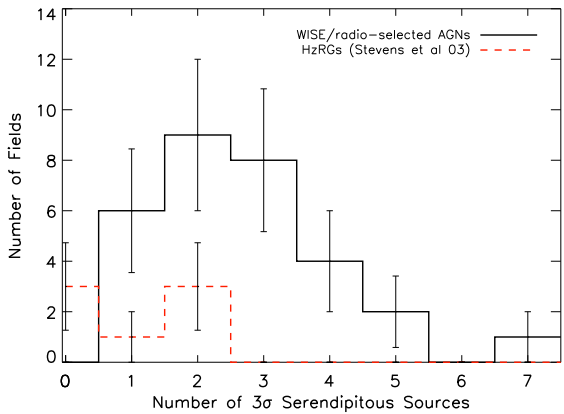}
\caption{The black solid line is the number of serendipitous sources per SCUBA-2 field for all 30 WISE/radio-selected AGNs, the red dotted line is the SMG serendipitous number counts around HzRGs (Figure 4 from \citealt{stevens03}). A higher number of SMGs are found around the WISE/radio-selected AGNs than HzRGs, see also Table 4. The typical noise level of our 30 maps is $\sim$ 2.1\,mJy/beam, which is comparable to $\sim$ 2.2\,mJy/beam from \citet{stevens03}.}
\label{histnum}
\end{figure}

\section{Discussion}

%The above results obtained for the WISE/radio-selected AGNs will be discussed by comparing their SEDs and luminosities with other galaxy populations. Next the WISE/radio-selected AGNs environments are investigated by comparing the serendipitous source number counts to other submm surveys. 

\subsection{WISE/radio-selected AGN SEDs and Luminosities}

The WISE/radio-selected AGN SEDs have bluer mid-IR sections than typical galaxies, while the SCUBA-2 observations show that they have relatively less submm emission than other galaxy SED templates: the detected WISE/radio-selected AGN targets' SCUBA-2 flux density is on average 3 times fainter than the bluest standard Polletta torus template. This leads to the suggestion that the WISE/radio-selected AGNs have less cold dust in the host galaxy and/or on the outer edge of the torus, and hence the torus could be denser, smaller and hotter than in the template. Alternatively, less submm emission could be due to an excess of mid-IR emission from the AGN as compared with the torus template \citep{wu12,tsai14}. The WISE/radio-selected AGN and Hot DOG SEDs are both dominated by mid-IR emission; however, the WISE/radio-selected AGNs are submm fainter and mid-IR bluer.

The luminosities of the four detected targets (with a mean luminosity of $L_{8\mu \textrm{m}-\textrm{SCUBA2}}$  = 2.9 $\times$ 10$^{13}$\,L$_{\odot}$) are higher than those of typical SMGs, which have $L_{\rm{IR}}$  = 8.5 $\times$ 10$^{12}$\,L$_{\odot}$ \citep{chapman05,kovacs06}, and DOGs, which have a mean luminosity $L_{\rm{IR}}$  = 9 $\times$ 10$^{12}$\,L$_{\odot}$ \citep{melbourne12}. High total IR luminosities were also found for the WISE/radio-selected AGNs from Lonsdale et al. (submitted) with a range of $L_{\rm{IR}}$ = 10$^{12} - 10^{13.55}$\,L$_{\odot}$. Hot DOGs had higher luminosities, with a mean luminosity of $L_{8\mu \textrm{m}-\textrm{SCUBA2}}$  = 5.3 $\times$ 10$^{13}$\,L$_{\odot}$ from \citet{jones14} and a mean luminosity of $L_{\rm{IR}}$ = 6.1 $\times$ 10$^{13}$\,L$_{\odot}$ from \citet{wu12}. However, WISE-selected AGNs could be biased towards being the mid-IR brightest and rarest galaxies, because they were selected on the grounds of their bright mid-IR flux.  Eight WISE-selected Lyman-alpha blobs (LABs) \citep{bridge13} were also found to be ultra-luminous galaxies from \textit{Herschel} data ($L_{\rm{IR}}$ = 2.3 $\times$ 10$^{13}$\,L$_{\odot}$), and included a wider range of mid-IR fluxes with no WISE colour selection cut. This indicates that galaxies with extremely red WISE colours have very luminous mid-IR properties. 

WISE/radio-selected AGN SEDs are dominated by mid-IR emission and are very luminous compared to other galaxy populations, suggesting the galaxies are very active from either star formation or AGN activity, and have large amounts of obscuring hot dust, that could be due to merging galaxies.

\subsection{WISE/radio-selected AGN Environments}

Comparing number counts of the serendipitous sources in the 30 WISE/radio-selected AGN fields with other submm surveys, implies there is an overdensity of SMGs in the 30 SCUBA-2 fields by factor of $\sim$ 4 - 6. This is consistent with finding WISE/radio-selected AGN in potentially overdense environments. This is an agreement with ALMA cycle-0 results of WISE/radio-selected AGNs \citep{silva14}, where 23 serendipitous SMG sources were detected in 17 out of 49 fields. This implies an overdensity factor of $\sim$ 10 compared to expected models on $\sim$ 20\,arcsec scales and unbiased population number counts \citep{silva14}. They report flux density limits with RMS noise level of $\sim 0.3 - 0.6$\,mJy beam$^{-1}$, which is lower than our RMS noise level of $\sim 1.8 - 3.7$\,mJy beam$^{-1}$). There is double the overdensity of SMG serendipitous sources in the ALMA fields than our SCUBA-2 fields when compared to previous submm surveys, this could be due to the multiplicity in SMGs \citep{karim13}. This is where multiple SMGs are found to be separated typically by $\sim$ 6\,arcsec and are resolved with high-resolution ALMA maps (1.5\,arcsec resolution), but are blended into one source with the resolution of single-dish surveys, where JCMT used in this paper has a resolution of 15\,arcsec at 850\,$\mu$m. \citet{karim13} suggested that the brightest SMGs (S$_{870\mu m}$ $\ge$ 12\,mJy) are significantly affected by multiplicity, but multiplicity may also influence fainter SMGs with submm flux densities of $\sim$ 4\,mJy. \citet{silva14} concluded that WISE/radio-selected AGNs reside in highly clustered environments, but further redshift data are needed to determine if they are in protocluster regions. 

Our SMG overdensity results are also consistent with previous submm observations of HzRGs. The number of serendipitous sources around seven HzRGs in SCUBA maps were twice the expected number from other blank-field surveys \citep{stevens03}: our SMG overdensity is higher compared with other blank-field surveys, by a greater factor of $\sim$ 4 - 6. There appears to be no correlation between the redshift of the WISE/radio-selected AGN and the number of serendipitous sources found in the fields around it however, the numbers here are only modest and more redshift data is needed to confirm this. 

\citet{stevens03} detected on average one serendipitous source per HzRG field, compared to our average of three serendipitous sources per WISE/radio-selected AGN SCUBA-2 field. The noise level in the \citet{stevens03} SCUBA fields was typically 2.1\,mJy/beam, which is comparable with the typical noise of 2.1\,mJy/beam in these SCUBA-2 fields. The results in Figure~\ref{histnum} and Table 4 agree that finding two 3\,$\sigma$ SMG companions per field is likely: however, no fields with no 3\,$\sigma$ serendipitous sources were found, and three 3\,$\sigma$ serendipitous sources are more likely per field. This suggests that these SCUBA-2 fields and WISE/radio-selected sources have a greater SMG overdensities compared with any previous surveys. 

The Hot DOG fields had a SMG overdensity factor of $\sim$ 2 - 3 compared with blank field submm surveys \citep{jones14} however, these WISE/radio-selected AGNs have an even higher overdensity. The K-correction effect applies at wavelengths longer than 250\,$\mu$m and the flux density from galaxies at redshifts $z > 1$ remains approximately constant with increasing redshift, up to $z \sim 10 - 20$ \citep{blain02}. Due to the K-correction effect, the serendipitous SMG detection is independent of redshift and so while the Hot DOGs have a typically higher redshift than the WISE/radio-selected AGNs, the higher overdensity of SMG sources around WISE/radio-selected AGNs is matched in luminosity of the  Hot DOGs companions. Four of the ten Hot DOGs from \citet{jones14} have radio emission; W1603$+$2745 S$_{1.4\rm{GHz}}$ = 2.8$\pm$0.5\,mJy, W1814$+$3412 S$_{1.4\rm{GHz}}$ = 3.4$\pm$0.5\,mJy, W1835$+$4355 S$_{1.4\rm{GHz}}$ = 3.9$\pm$0.4\,mJy, W2216$+$0723 S$_{1.4\rm{GHz}}$ = 5.9$\pm$0.4\,mJy. However, the typical radio emission for Hot DOGs is 4\,mJy, which is lower than WISE/radio-selected AGN value of 59\,mJy. It suggests that the WISE/radio-selected galaxies which have radio emission are found in more overdense regions of the sky, compared to Hot DOGs with no or low radio emission. This is further evidence that AGN with radio-intermediate/loud emission reside in denser environments than radio-quiet AGN. This is not thought to be a selection bias, but rather indicates that RLAGN are located in more massive haloes \citep{yates89,hill91,best98,roche98,best00,donoso10,hatch14}. Our SMG overdensity could indicate that WISE/radio-selected AGNs are signposts of protocluster regions.

None of the serendipitous SMG sources in the WISE/radio-selected AGN and Hot DOG fields were detected in the NVSS or FIRST catalogues. However, 48\,$\%$ and 59\,$\%$ of serendipitous SMG sources in the WISE/radio-selected AGN and Hot DOG fields, respectively, had counterparts in the AllWISE Source Catalog. All serendipitous sources detected around WISE/radio-selected AGNs were detected in the W1 band, 74\,$\%$ in the W2 band, 10\,$\%$ in the W3 band, and 0\,$\%$ in the W4 band. All serendipitous sources detected around Hot DOGs were detected in the W1 band, 90\,$\%$ in the W2 band, 20\,$\%$ in the W3 band, and 0\,$\%$ in the W4 band. Figure~\ref{hist85022} shows the W4 flux density versus SCUBA-2 850\,$\mu$m flux density of the 39 serendipitous sources detected around the WISE/radio-selected AGNs that have WISE data, compared with WISE/radio-selected AGNs and Hot DOGs. From Figure~\ref{hist85022} it can be seen that the serendipitous sources are less bright in W4 and therefore, not as red in the mid-IR than WISE/radio-selected AGNs and Hot DOGs. This implies they are normal SMGs, that have $S_{850\mu \rm{m}}$ $> 2mJy$, and are high-redshift galaxies with high IR luminosities believed to be from starburst activity, but are faint in optical and near-IR wavelengths \citep{ivison98,eales99,smail00,barger00,chapman01,blain02,ivison02,ivison04,pope06}. Most of the serendipitous SMG sources will be at a similar redshift (1 $\le$ z $\le$ 3) to the WISE/radio-selected AGNs \citep{karim13}, but some will be inevitably unassociated in redshift. 

These WISE/radio-selected AGNs appear to be very powerful AGN that have more mid-IR emission, and mid-IR opacity than AGN in standard galaxy templates. Therefore, the WISE/radio-selected AGNs might be experiencing the most powerful AGN feedback possible and could be an obscured AGN-dominated short evolutionary phase of merging galaxies, for example the Hot DOG target W1814$+$3412 has three components of the same redshift and within 50kpc of each other \citep{eisenhardt12}. The WISE/radio-selected AGNs also appear to reside in intriguing many-arcmin-scale overdensities of very luminous, dusty sources. 

WISE/radio-selected AGNs have lower submm flux densities, higher radio emission, lower redshifts and denser environments than Hot DOGs. These could be due to Hot DOGs having a higher typical redshift, and were selected to be mid-IR redder. WISE/radio-selected AGNs and Hot DOGs appear to be consistent with an AGN-dominated population of galaxies, that could be a transient phase of the major merger theory, but starburst activity cannot be ruled out. The WISE/radio-selected AGNs appear to be signposts of overdense regions of active, dusty and luminous galaxies. The next chapter will further discuss the Hot DOGs and WISE/radio-selected AGNs compared with each other, and their serendipitous SMG sources. 

\section{Summary}

The results from SCUBA-2 850\,$\mu$m observations of 30 WISE/radio-selected, high-redshift, luminous, dusty AGNs are:

\begin{itemize}

\item The 30 WISE/radio-selected AGNs have SEDs that are not well fitted by the current AGN templates (see Figure~\ref{sedall}); the best fitting is the Polletta torus \citep{polletta07} template. 

\item The detected WISE/radio-selected AGNs have less cold dust than the Polletta torus template, which could be because there is less cold dust in the host galaxy, and/or the outer scale of the AGN torus in the WISE/radio-selected AGNs are smaller. Alternatively there could be more intense mid-IR emission from hotter inner regions \citep{wu12}. 

\item Despite being observed over a wide redshift range, the 10 WISE/radio-selected AGNs with known redshift data show uniform submm to mid-IR ratios. The highest redshift, most luminous targets, could thus have hotter dust temperatures than assumed in the templates. However, the number of targets with known redshifts is currently only modest (10) and the selection of the targets is sensitive to redshift, owing to very red WISE colours.

\item The detected WISE/radio-selected AGNs have high IR luminosities, $L_{8\mu \textrm{m}-\textrm{SCUBA2}}$ $\ge$ 10$^{13}$ L$_{\odot}$, confirming they are HyLIRGs. These are conservative values as any pronounced peak of the SED would increase these further. The undetected WISE/radio-selected AGNs have upper limit luminosities that are consistent with LIRGs.

\item The luminosity of an underlying extended star-forming galaxy cannot exceed a luminosity $\sim$ 4\,$\%$ (for a cool spiral galaxy template) or $\sim$ 10\,$\%$ (for a warmer ULIRG-like galaxy template) as compared with the submm-detected typical WISE/radio-selected AGNs luminosity. Our SCUBA-2 observations confirm that WISE/radio-selected AGNs are a mid-IR dominated population.

\item When comparing the submm galaxy counts of the 30 1.5-arcmin-radius SCUBA-2 maps observed here to blank-field surveys, there is an overdensity of SMGs on this scale by a factor 4 - 6, but no evidence for any angular clustering within these fields. 

\item There is a SMG overdensity of order $\sim$ 2 when comparing WISE/radio-selected AGNs to Hot DOGs, and suggests that WISE/radio-selected AGNs are signposts of overdense regions of active, luminous and dusty galaxies in the sky.

\item WISE/radio-selected AGNs and Hot DOGs appear to be consistent with an AGN-dominated population of galaxies, but starburst activity cannot be ruled out, and could be a transient phase of the major merger theory.

%\item Future work will be to find the spectroscopic redshifts for the rest of the sample, as well as more submm observations to increase the sample size, and far-IR observations to be able to accurately define the peak of the SED.

\end{itemize}

\section{Acknowledgements}

The authors would like to thank the anonymous referee for his/her comments and suggestions, which have greatly improved this paper.

S. F. Jones gratefully acknowledges support from the University of Leicester Physics \& Astronomy Department. This publication makes use of data products from the \textit{Wide-field Infrared Survey Explorer}, which is a joint project of the University of California, Los Angeles, and the Jet Propulsion Laboratory/California Institute of Technology, funded by the National Aeronautics and Space Administration.

The James Clerk Maxwell Telescope has historically been operated by the Joint Astronomy Centre on behalf of the Science and Technology Facilities Council of the United Kingdom, the National Research Council of Canada and the Netherlands Organisation for Scientific Research. Additional funds for the construction of SCUBA-2 were provided by the Canada Foundation for Innovation. The programme IDs under which the data were obtained were M12BU07 and M13BU02.

RJA was supported by Gemini-CONICYT grant number 32120009.

\bibliographystyle{mn2e}
\bibliography{ref}

\label{lastpage}
\end{document}